\documentclass[preprints,article,accept,pdftex,oneauthors]{Definitions/mdpi} 
\firstpage{1} 
\pubvolume{1}
\issuenum{1}
\articlenumber{0}
\pubyear{2025}
\copyrightyear{2025}
\datereceived{ } 
\daterevised{ } 
\dateaccepted{ } 
\datepublished{ } 
\hreflink{https://doi.org/} 

\usepackage{siunitx}
\newcommand{\sech}{\operatorname{sech}}
\usepackage{multirow}

\Title{Cosmic-Ray Boosted Diffuse Supernova Neutrinos}

\TitleCitation{Cosmic-Ray Boosted Diffuse Supernova Neutrinos}

\Author{Alexander Sandrock $^{1}$\orcidA{}}

\AuthorNames{Alexander Sandrock}

\isAPAStyle{%
       \AuthorCitation{Sandrock, A.}
         }{%
        \isChicagoStyle{%
        \AuthorCitation{Sandrock, Alexander.}
        }{
        \AuthorCitation{Sandrock, A.}
        }
}

\address[1]{%
$^{1}$ \quad University of Wuppertal, Germany; asandrock@uni-wuppertal.de
}

\corres{Correspondence: asandrock@uni-wuppertal.de
}

\abstract{The subject of boosted fluxes of dark matter or cosmic relic neutrinos
via scattering on cosmic rays has received considerable attention recently. This
article investigates the boosted neutrino flux from scattering of cosmic rays
and the so-far undetected diffuse supernova neutrino background, taking into
account both galactic and extragalactic cosmic rays. The calculated flux is many
orders of magnitude smaller than either the galactic diffuse neutrino emission,
the extragalactic astrophysical flux measured by IceCube, or the cosmogenic
neutrino flux expected at the highest energies.}


\begin{document}

\section{Introduction}
Core collapse supernov\ae{} emit many orders of magnitude more energy in
neutrinos than in photons and the most recent core collapse supernova in the
near neighborhood of the Milky Way, SN1987A, was the first extraterrestrial
neutrino source ever detected apart from the Sun. For a long time, a diffuse
isotropic background of neutrinos from the integrated emission of many core
collapse supernov\ae{} has been predicted, but so far it has escaped detection.
Theoretical predictions vary by a factor of 2 to 5, and existing experimental
limits are about an order of magnitude higher than predictions
\cite{super_k_dsnb,juno_dsnb}.

Recently, calculations of the flux of relic neutrinos upscattered by elastic
scattering on cosmic ray particles to ultra-high energies have regained
interest \cite{boosted_jiajun,de_marchi}. The comparison of the computed fluxes
to experimental upper limits on ultra-high energy neutrino fluxes have allowed
the authors to determine strong upper limits on the density of relic neutrinos.

This article calculates the expected neutrino flux, resulting from scattering
of the diffuse supernova neutrino background on cosmic rays in order to
investigate the possibility of an indirect detection of this neutrino background
by observation of high-energy neutrinos. There are two contributions with
different angular distribution. On the one hand, galactic cosmic rays can
upscatter the diffuse supernova neutrinos; the resulting neutrino flux will
then be concentrated around the galactic plane, similar to the diffuse galactic
neutrino flux from hadronuclear interactions of cosmic rays with the interstellar
medium \cite{cringe}. On the other hand, extragalactic cosmic rays can
upscatter the diffuse supernova neutrino background, resulting in an isotropic
flux of ultrahigh energy neutrinos, similar to the cosmogenic neutrino flux
expected from photohadronic interactions of ultra-high energy cosmic rays with
the cosmic neutrino background \cite{berat_cosmogenic}, which has so far not
been detected \cite{ehe_prl,ehe_moriond,auger}.

\section{Diffuse Supernova Neutrino Background}
\begin{figure}
  \begin{center}
  \includegraphics[width=0.7\textwidth]{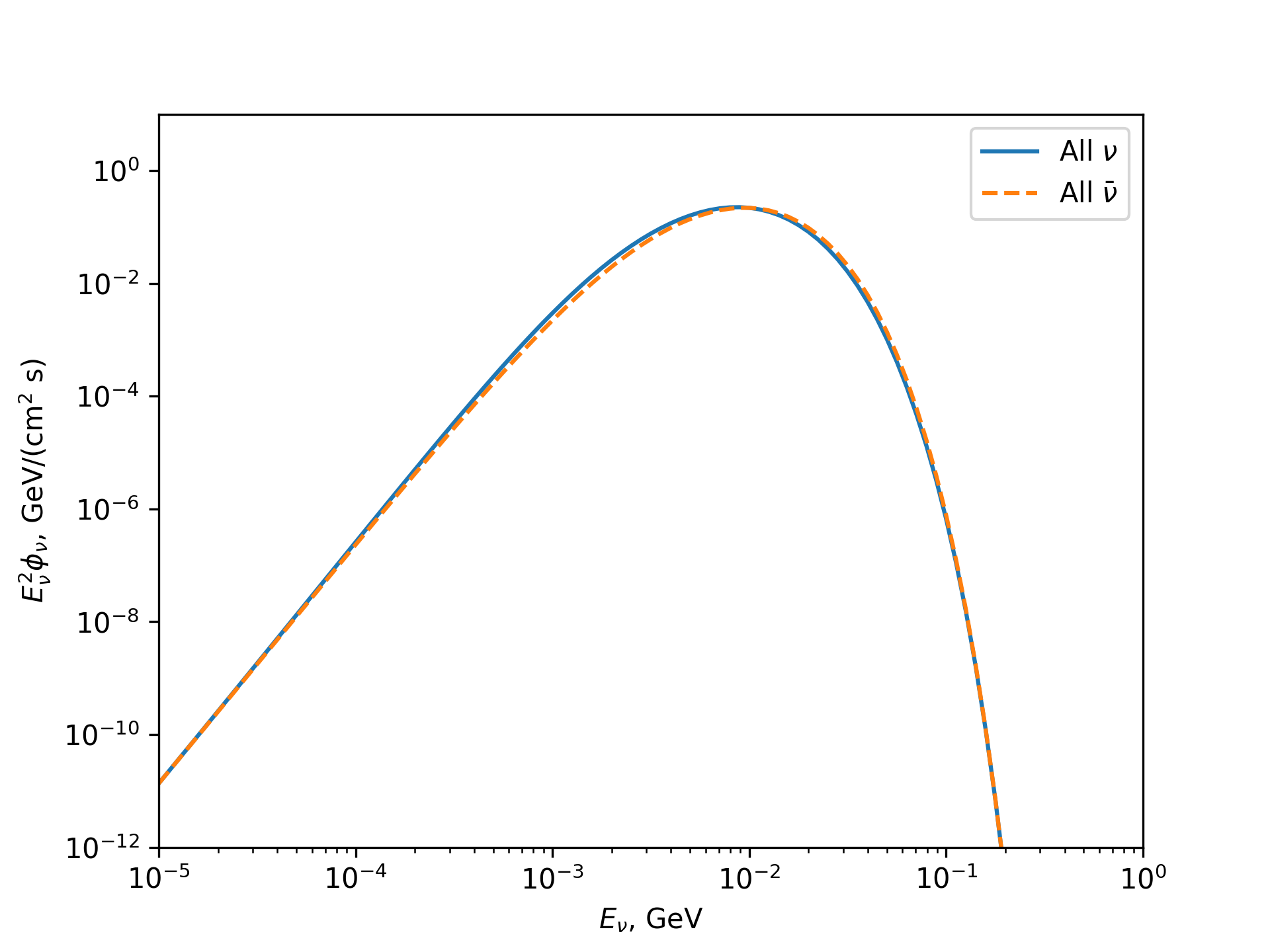}
  \end{center}
  \caption{Fiducial spectrum of the diffuse supernova neutrino background from
    \cite{dsnb_flux} based on simulations by \cite{garching_sim}. The spectrum
    is weighted with the square of the neutrino energy.}
  \label{fig:DSnuB}
\end{figure}

For this calculation, the flux prediction by \cite{dsnb_flux} is used, who
present a convenient parametrization of SN~simulations by \cite{garching_sim}.
Since the neutral-current neutrino cross-sections do not depend on the neutrino
flavour, only the total sum of neutrinos and antineutrinos is needed.
The flux prediction is given by the integral
\begin{equation}
  \phi_{\nu_l} = c \int_{8 M_\odot}^{125 M_\odot} dM \int_0^5 dz\ (1 + z)
  \left|\frac{dt_c}{dz}\right| R_\text{SN}(z, M) Y_{\nu_l}(E_\nu (1 + z), M),
\end{equation}
where $Y_{\nu_l}$ refers to the parametrized neutrino yield,
\begin{equation}
  R_\text{SN}(z, M) = \dot\rho_*(z) \frac{\phi(M)\, dM}
  {\int_{0.5 M_\odot}^{125 M_\odot} \phi(M) M\,dM}
\end{equation}
is the supernova rate, determined from star formation history $\dot\rho_*$ and
initial mass function $\phi$ with $M$ the stellar mass, and
\begin{equation}
  \left|\frac{dt_c}{dz}\right| = \left[H_0 (1 + z) \sqrt{\Omega_\Lambda
  + (1 + z)^3 \Omega_m}\right]^{-1}.
\end{equation}
describes the expansion history of the universe with redshift $z$.
The fiducial spectrum of the diffuse supernova neutrino background is shown in
figure~\ref{fig:DSnuB}.

\section{Cosmic Ray Flux}
\subsection{Galactic Cosmic Ray Flux}
\begin{figure}
  \begin{center}
  \includegraphics[width=0.7\textwidth]{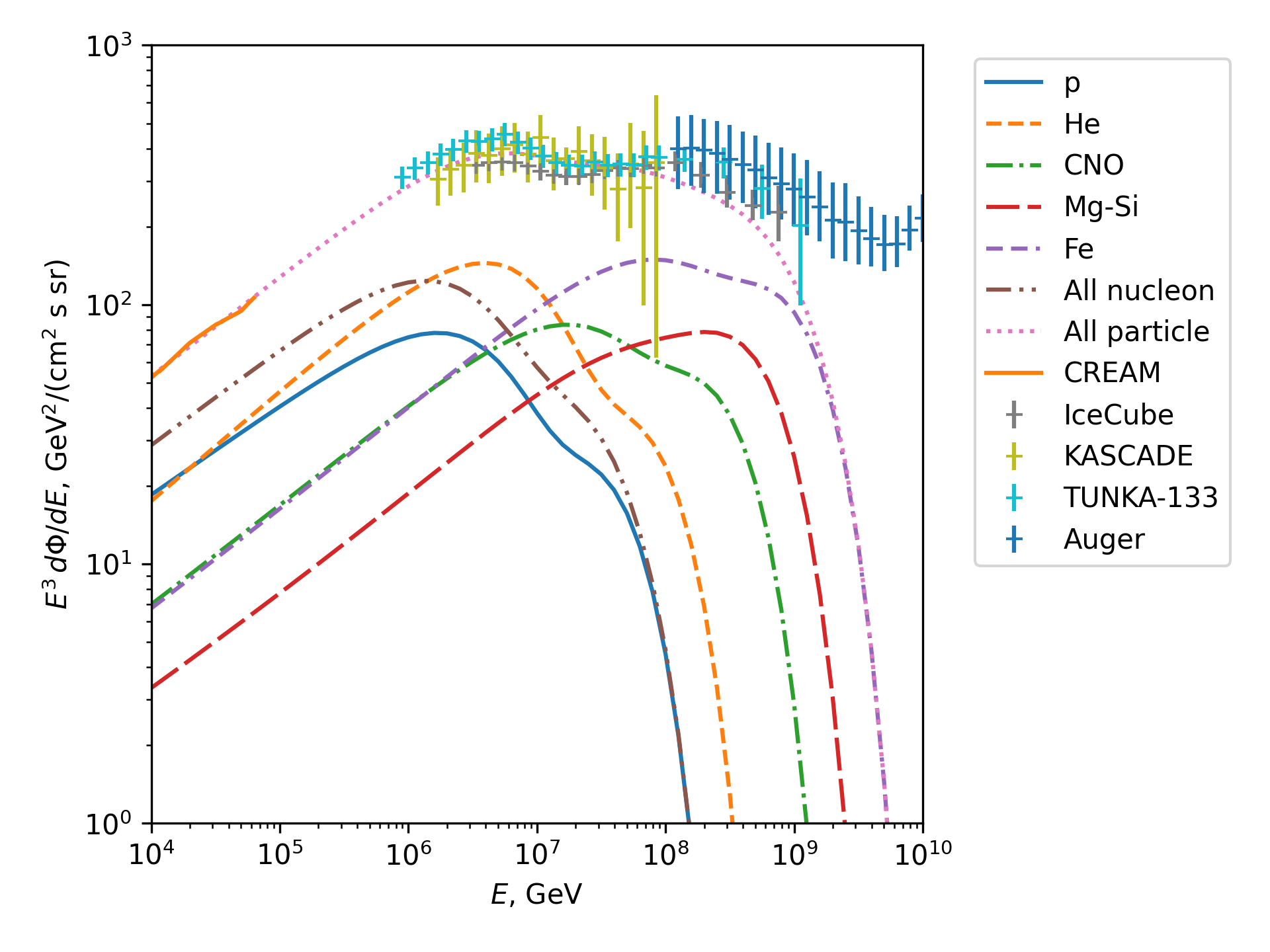}
  \end{center}
  \caption{Galactic cosmic ray spectrum according to the mixed composition model from
  \cite{gaisser_H3a}, weighted with the third power of the energy per particle.
  The third generation is set to zero, since the extragalactic cosmic rays are
  treated differently. Measurements of the all-particls cosmic ray flux from
  CREAM, IceCube, KASCADE, TUNKA, and the Pierre Auger observatory
  are shown for comparison (taken from the compilation in
  \cite{2023EPJC...83..971M,The_CR_spectrum}).}
  \label{fig:gaisser_H3a}
\end{figure}

For the galactic cosmic ray flux, the parametrization by \cite{gaisser_H3a} is
adopted, which describes the cosmic ray spectrum as the sum of several
exponentially cutoff power-law spectra in rigidity $R = E/Ze$
(cf.~fig.~\ref{fig:gaisser_H3a}). The flux for a nucleus $i$ is then given by
\begin{equation}
  \frac{d\Phi_i}{dE_i} = \sum_{j = 1}^3 a_{ij} E_i^{-\gamma_j - 1}
  \exp\left(-\frac{E_i}{Z_i R_j^\text{cut}}\right)
\end{equation}
corresponding to three populations of cosmic rays with cutoff rigidity
$R_j^\text{cut}$ and spectral index $\gamma_j$ common for each population
(except for the spectral index of the first population, which is different for
each species, based on direct measurements).
The third generation of this parametrization corresponds to the extragalactic
cosmic ray flux. This is treated differently (see following section), therefore
the normalization of the third generation is set to zero.

\subsection{Extragalactic Cosmic Ray Flux}
The extragalactic cosmic ray flux is determined using simulation results of
the UHECR propagation code \textit{PriNCe} \cite{heinze}. Since the sources
of UHECRs are so far unknown, we assume three possible source density scalings
in this work, namely the star formation rate
\cite{star_formation_rate}
\begin{equation}
  N_\text{SFR}(z) = \frac{1 + a_2 z/a_1}{1 + (z/a_3)^{a_4}}
\end{equation}
with $a_1 = \num{0.0170}, a_2 = \num{0.13}, a_3 = \num{3.3}, a_4 = \num{5.3}$,
the quasar rate \cite{quasar_rate}
\begin{equation}
  N_\text{QSO}(z) = \exp \left(a_1 z - a_2 z^2 + a_3 z^3 - a_4 z^4\right)
\end{equation}
with $a_1 = \num{2.704}, a_2 = \num{1.145}, a_3 = \num{0.1796},
 a_4 = \num{0.01019}$, and the gamma-ray burst rate
\begin{equation}
  N_\text{GRB}(z) = N_\text{SFR}(z) (1 + z)^\delta
\end{equation}
with $\delta = \num{1.26}$. In each case, the normalization of the source
density is chosen such that $N_i(z = 0) = 1$. This does not necessarily imply
that star forming galaxies, quasars or gamma-ray bursts are the sources of
ultra-high energy cosmic rays; the assumption is only that they scale with
redshift in the same way.

The spectrum of the injected cosmic rays is assumed to follow the spectral form
\cite{auger_combined_fit}
\begin{equation}
  J_A(E) = N_\text{evol}(z) f_A \left(\frac{E}{\SI{e9}{GeV}}\right)^{-\gamma}
  \begin{cases}
    1 & E < Z R_\text{max}, \\
    \exp\left(1 - \frac{E}{Z R_\text{max}}\right) & E > Z R_\text{max},
  \end{cases}
\end{equation}
where $A, Z$ are the mass and charge number of the injected nucleus, $E$
its energy, $R_\text{max}$ the maximum rigidity, $\gamma$ the spectral index,
$N_\text{evol}(z) \in \{N_\text{SFR}, N_\text{QSO}, N_\text{GRB}\}$ the redshift
evolution of the source density, and $f_A$ the normalization.
By a fit to measurements from the Pierre Auger Observatory \cite{auger2019},
the spectral index, maximum rigidity and normalizations of the injected source
spectrum is determined (see Table~\ref{tab:extragalactic}).
\begin{table}
\caption{Fit parameters of the UHECR fit for the specified source distribution}
\label{tab:extragalactic}
\begin{tabularx}{\textwidth}{l|rrr}
\toprule
Source density scaling & Star formation rate & Quasar rate & Gamma-ray burst rate \\
\\
\midrule
$R_\text{max}$, \si{GV} & $10^{9.25}$ & $10^{9.20}$ & $10^{9.25}$ \\
$\gamma$ & \num{-0.8} & \num{-1.0} & \num{-0.8} \\
$f_\text{H}$, \si{GeV^{-1} cm^{-3} s^{-1}}  & \num{1.0e-45} & \num{2.2e-45} & \num{1.7e-46} \\
$f_\text{He}$, \si{GeV^{-1} cm^{-3} s^{-1}} & \num{4.5e-46} & \num{7.7e-46} & \num{4.6e-47} \\
$f_\text{N}$, \si{GeV^{-1} cm^{-3} s^{-1}}  & \num{6.2e-47} & \num{6.1e-47} & \num{5.5e-47} \\
$f_\text{Si}$, \si{GeV^{-1} cm^{-3} s^{-1}} & \num{4.1e-48} & \num{2.3e-48} & \num{3.9e-48} \\
$f_\text{Fe}$, \si{GeV^{-1} cm^{-3} s^{-1}} & \num{9.0e-50} & \num{6.6e-50} & \num{9.5e-50} \\
\bottomrule
\end{tabularx}
\end{table}
\begin{figure}
\centering
\includegraphics[width=0.7\textwidth]{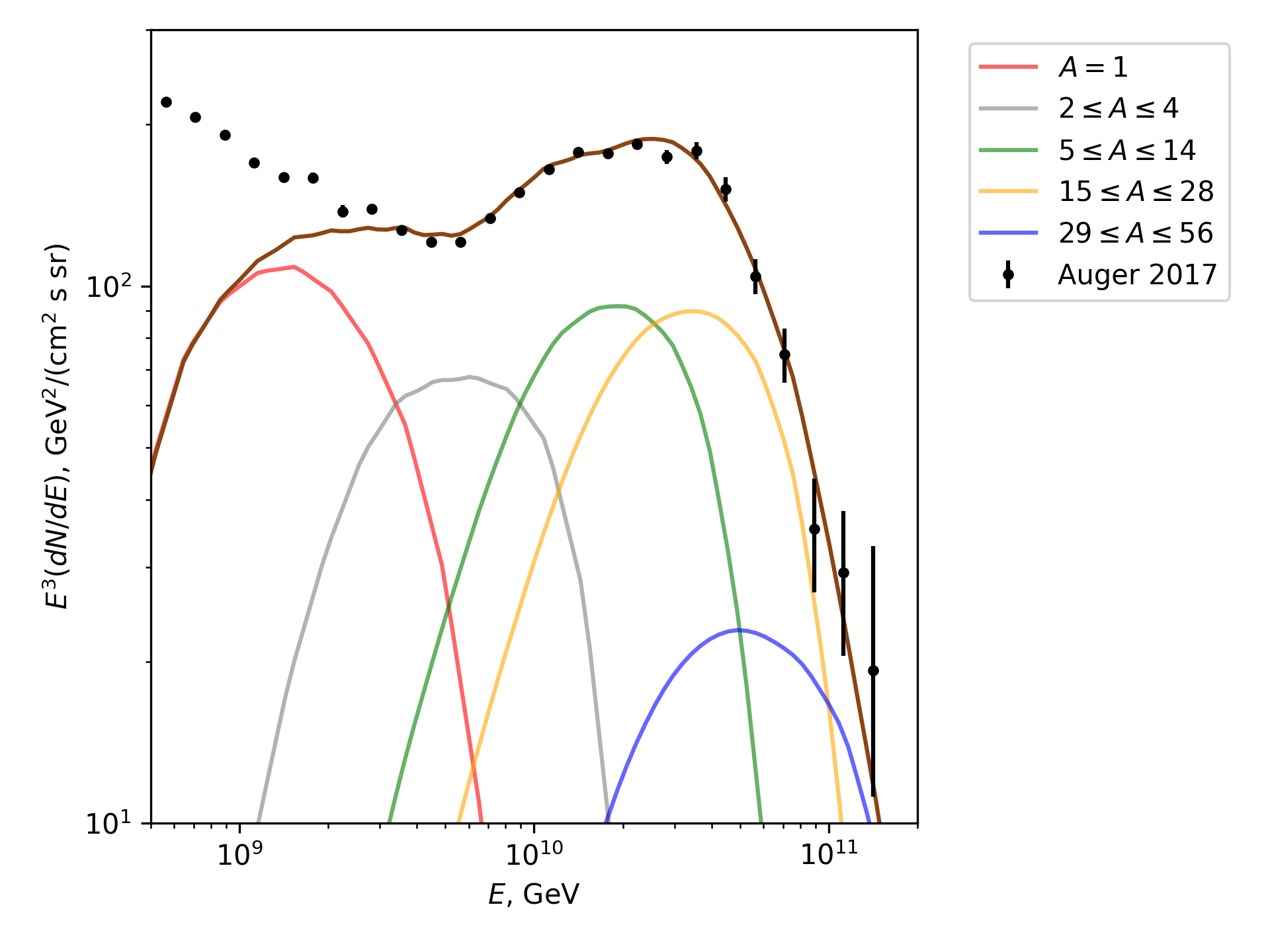}
\caption{Extragalactic cosmic ray flux fit, assuming a source density scaling
as the star formation rate. The fit is to the UHECR measurement data from
\cite{auger2019} above a lower energy of \SI{6e9}{GeV}.}
\label{fig:extragalactic}
\end{figure}
The cosmological parameters used in this fit are $H_0 = \SI{70.5}{km/s/Mpc}$ for
the Hubble constant, and $\Omega_m = \num{0.27}, \Omega_\Lambda = \num{0.73}$.
The photodisintegration cross-section used is the TALYS parametrization \cite{talys}.
The fitted spectrum is shown in Figure~\ref{fig:extragalactic} for density scaling
according to the star formation rate, and in Figure~\ref{fig:QSO_UHECR},
\ref{fig:GRB_UHECR} for the quasar and gamma-ray burst rate, respectively.

\section{Neutrino-Nucleon and Neutrino-Nucleus Scattering Cross-Sections}
\label{app:xsec}
For the elastic scattering of neutrinos on nucleons, the parametrization of
\cite{ahrens}
\begin{equation}
\begin{split}
  \frac{d\sigma}{dQ^2} &= \frac{G_\text{F}^2 M_p^2}{8 \pi (E_\nu^*)^2}
    \left(A(Q^2) \pm \frac{s - u}{M_p^2} B(Q^2) + \frac{(s - u)^2}{M_p^4} C(Q^2)
    \right), \\
  s - u &= 4 M_p (E_\nu^*) - Q^2,
\end{split}
\label{eq:ahrens}
\end{equation}
is used. The $\pm$ in front of $B(Q^2)$ refers to $+$ for neutrinos
and $-$ for antineutrinos. The coefficients $A(Q^2), B(Q^2), C(Q^2)$ are
functions of the vector and axial vector form factors of the proton
(see Appendix~\ref{app:neutrino}). The neutrino energy in the rest frame of the
nucleon is denoted by $E_\nu^*$, the momentum transfer squared by $Q^2$, and the
proton mass by $M_p$; $s, u$ are Mandelstam variables of two-particle scattering
\cite{Mandelstam}.

The coherent elastic neutrino-nucleus scattering cross-section can be written
as \cite{formaggio}
\begin{equation}
  \frac{d\sigma}{dQ^2} = \frac{G_\text{F}^2}{8 \pi} Q_\text{W}^2
  \left(1 - \frac{Q^2}{4 (E_\nu^*)^2}\right) F(Q^2)^2
\end{equation}
with $F(Q^2)$ the nuclear formfactor \cite{klein} and
\begin{equation}
  Q_W = (A - Z) - Z (1 - 4 \sin^2 \theta_\text{W})
\end{equation}
the weak charge of the nucleus. The incoherent elastic neutrino-nucleus
scattering cross-section can be written as
\begin{equation}
  \frac{d\sigma}{dQ^2} = \left. \frac{d\sigma}{dQ^2} \right|_\text{elastic
  scattering} A [1 - F(Q^2)^2].
\end{equation}

The differential cross-section of deep inelastic scattering is given by
\begin{equation}
  \frac{d^2\sigma}{dy\ dQ^2} = \frac{G_\text{F}^2 M_Z^4}{4 \pi y (Q^2 + M_Z^2)^2}
  \{[1 + (1 - y)^2] F_2(x, Q^2) - y^2 F_L(x, Q^2)\},
\end{equation}
with $y$ the inelasticity and $x$ the Björken scaling variable. In this work,
the parametrization of the structure function $F_2$ from \cite{block} is used,
in combination with the parametrization of the structure function ratio
$R = F_L/(F_2 - F_L)$ from \cite{abe}. 

\section{Galactic Boosted Neutrino Flux}
\subsection{Elastic scattering}
The rate of upscattered neutrinos per volume, energy and solid angle can be
written as
\begin{equation}
\begin{split}
  \dot n_s (\epsilon_s, \Omega_s) &= \frac{dn_s}{d\epsilon_s\ d\Omega_s} \\
  &= c \int_0^\infty d\epsilon_\nu \oint d\Omega_\nu
    \int_1^\infty d\Gamma_\text{CR} \oint d\Omega_\text{CR}
    (1 - \beta \cos \psi) n_\nu n_\text{CR}
    \frac{d\sigma}{d\epsilon_s\ d\Omega_s}
\end{split}
\end{equation}
with $\cos \psi = \cos \theta_\nu \cos \theta_\text{CR} + \sin \theta_\nu 
\sin \theta_\text{CR} \cos (\phi_\nu - \phi_\text{CR})$ and the DSN$\nu$B
density $n_\nu = dn_\nu/(d\epsilon_\nu\ d\Omega_\nu)$, and the cosmic ray
density $n_\text{CR} = dn_\text{CR}/(d\Gamma_\text{CR}\ d\Omega_\text{CR})$.
The polar angles $\Omega = (\theta, \phi)$, energies $\epsilon$, and
Lorentz factor $\Gamma$  refer to the upscattered neutrino (subscript $s$), DSN$\nu$B
neutrino (subscript $\nu$), and the cosmic ray particle (subscript CR). From
kinematic considerations, the head-on approximation
$\Omega_s = \Omega_\text{CR}$ and relative velocity $\beta \approx 1$
are appropriate for this problem because $\Gamma_\text{CR} \gg 1$.

The energy of the neutrino in the nucleus (nucleon) rest frame is given by
\begin{equation}
  \epsilon_\nu^* = \epsilon_\nu \Gamma_\text{CR} (1 - \beta \cos \psi)
\end{equation}
and the squared momentum transfer by
\begin{equation}
  Q^2 = 2 \epsilon_\nu \epsilon_s (1 - \cos \psi).
\end{equation}
Using this together with the approximations above, the expression
\begin{equation}
\begin{split}
  \dot n_s(\epsilon_s, \Omega_s) &= c \int_0^\infty d\epsilon_\nu
    \oint d\Omega_\nu \int_1^\infty d\Gamma_\text{CR}
    2 \epsilon_\nu n_\nu n_\text{CR}|_{\Omega_\text{CR} = \Omega_s} \\
  &\times \left. \frac{d\sigma}{dQ^2} \right|^{\epsilon^*_\nu = \epsilon_\nu
      \Gamma_\text{CR} (1 - \cos \psi)}_{Q^2 = 2 \epsilon_\nu \epsilon_s
      (1 - \cos \psi)}
\end{split}
\end{equation}
is obtained. Since both the diffuse neutrino background and the cosmic-ray flux
are isotropic to very good approximation, we can choose the polar axis of the
coordinate system of $\Omega_\nu$ to coincide with the cosmic-ray direction
$\Omega_\text{CR}$. Then $\cos\psi = \cos\theta$ and the integral over the
azimuthal angle becomes trivial. The integral over the polar angle can be
expressed as an integral over $Q^2$, so that we are left with
\begin{equation}
  \dot n_s = \frac{\pi c}{2 \epsilon_s^3}
    \int_0^\infty \frac{d\epsilon_\nu}{\epsilon_\nu^2}
    \int_1^\infty d\Gamma_\text{CR}\ n_\nu n_\text{CR}
    \int_0^{4 \epsilon_s \epsilon_\nu} dQ^2\ Q^4 \frac{d\sigma}{dQ^2}.
\end{equation}

\subsection{Deep inelastic scattering}
While in elastic scattering the final state is determined by a single variable,
in the inelastic case two variables are needed, requiring a separate
calculation. In the head-on approximation, the inelasticity $y$ is connected to
the final energy $\epsilon_s$ in the galactic rest frame as
\begin{equation}
  \epsilon_s = 2 (1 - y) \epsilon_\nu \Gamma_\text{CR}^2 (1 - \beta \cos \psi),
\end{equation}
so that the cross-section differential in $\epsilon_s$ is given by
\begin{equation}
  \frac{d\sigma}{d\epsilon_s} = \frac{1}{2 \epsilon_\nu \Gamma_\text{CR}^2
    (1 - \beta \cos \psi)} \left. \frac{d\sigma}{dy} \right|_{
    y = 1 - \epsilon_s/[2 \epsilon_\nu \Gamma_\text{CR}^2 (1 - \beta \cos\psi)]}^{
    \epsilon_\nu^* = \epsilon_\nu \Gamma_\text{CR} (1 - \beta \cos \psi)}.
\end{equation}
In analogy with the above calculation,
\begin{equation}
\begin{split}
  \dot n_s (\epsilon_s, \Omega_s) &= c \int_0^\infty d\epsilon_\nu
    \oint d\Omega_\nu \int_1^\infty d\Gamma_\text{CR} \oint d\Omega_\text{CR}\ 
    (1 - \beta \cos \psi) n_\nu n_\text{CR} \frac{d^2 \sigma}{d\epsilon_s\ 
    d\Omega_s} \\
  &\simeq c \int_0^\infty d\epsilon_\nu \oint d\Omega_\nu
    \int_1^\infty d\Gamma_\text{CR}\ (1 - \beta \cos \psi) n_\nu n_\text{CR}
    \frac{d\sigma}{d\epsilon_s} \\
  &= c \int_0^\infty d\epsilon_\nu \oint d\Omega_\nu \int_1^\infty d\Gamma_\text{CR}
    \frac{n_\nu n_\text{CR}}{2 \epsilon_\nu \Gamma_\text{CR}^2}
    \left. \frac{d\sigma}{dy} \right|_{
    y = 1 - \epsilon_s/[2 \epsilon_\nu \Gamma_\text{CR}^2 (1 - \beta \cos\psi)]}^{
    \epsilon_\nu^* = \epsilon_\nu \Gamma_\text{CR} (1 - \beta \cos \psi)} \\
  &= \pi c \int_0^\infty d\epsilon_\nu \frac{n_\nu}{\epsilon_\nu}
    \int_1^\infty d\Gamma_\text{CR} \frac{n_\text{CR}}{\Gamma_\text{CR}^2}
    \int_0^{2 \epsilon_\nu \Gamma_\text{CR}} \frac{d\epsilon_\nu^*}
    {\epsilon_\nu \Gamma_\text{CR}}
    \int\limits_0^{2 m_p (y \epsilon_\nu^* - m_\pi) - m_\pi^2} dQ^2\ \frac{d^2\sigma}{dy\ dQ^2}.
\end{split}
\end{equation}
Since the minimal energy in the nucleon rest frame for this process is given by
$\epsilon^*_\text{min} = m_\pi + m_\pi^2/2 m_p$, the integration region is more
constrained compared to the elastic scattering case, leading to
\begin{equation}
\begin{split}
  \dot n_s (\epsilon_s, \Omega_s) &= \pi c \int_0^\infty d\epsilon_\nu
  \int_{\epsilon^*_\text{min}/2 \epsilon_\nu}^\infty d\Gamma_\text{CR}
  \int_{\epsilon^*_\text{min}}^{2 \epsilon_\nu \Gamma_\text{CR}} d\epsilon^*
  \int\limits_0^{2 m_p (\epsilon_\nu^* - \epsilon_s/2 \Gamma_\text{CR} - m_\pi) - m_\pi^2} dQ^2
    \\ 
  &\times
  \frac{n_\nu n_\text{CR}}{\epsilon_\nu^2 \Gamma_\text{CR}^3} \frac{d^2\sigma}{dy\ dQ^2}.
\end{split}
\end{equation}

\subsection{Local neutrino emission rate}
The emission rate varies as a function of the local cosmic-ray spectrum, while
the diffuse supernova neutrino flux can be assumed to be constant over the scale
of the galaxy. 

\begin{figure}
  \begin{center}
  \includegraphics[width=0.7\textwidth]{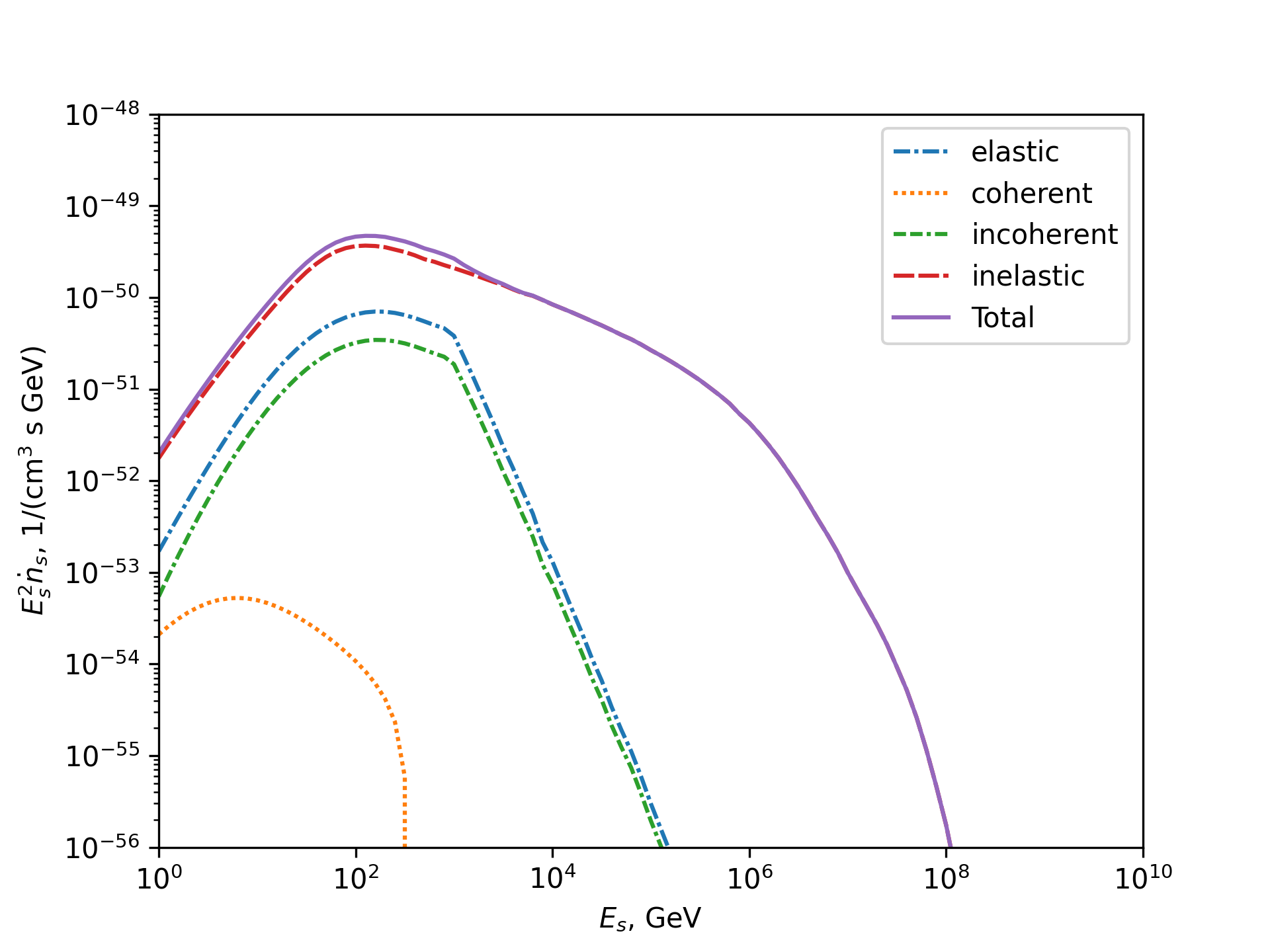}
  \end{center}
  \caption{Emission rate per volume $\dot n_s$ at earth for elastic neutrino-nucleon
    scattering, coherent and incoherent neutrino-nucleus scattering, and deep-inelastic
    neutrino-nucleus scattering of diffuse supernova neutrinos on galactic cosmic rays.}
  \label{fig:emission_rate}
\end{figure}
The numerical results are shown in Figure~\ref{fig:emission_rate}. The dominant
contribution to the emission rate is due to inelastic scattering, with elastic
scattering on a proton having a small, yet non-negligible impact at energies up
to about \SI{1}{TeV}. For energies below a few hundred \si{GeV}, the spectral
index is about \num{-1}, while for larger energies the emission rate roughly
follows the cosmic ray spectrum with a spectral index of \num{-2.4} and
steepening at energies higher than a \si{PeV}.

\subsection{Intensity of boosted neutrinos from galactic cosmic rays}
The intensity of the boosted flux follows the spatial dependence of the cosmic-ray
flux, since the diffuse supernova neutrino background is homogeneous on the scale
of the galaxy. Therefore, the intensity map follows by a line-of-sight integral as
\begin{equation}
  I(\epsilon_s, \Omega) = \dot n_s|_\odot \int_0^\infty f(\vec x_\odot + \tau \hat \Omega) d\tau,
\end{equation}
where $\hat \Omega$ denotes a unit vector in the direction $\Omega$. The
spectral shape of the cosmic-ray spectrum is identical everywhere, if
the model describing the Galactic cosmic rays assumes a steady state,
the source spectra are the same everywhere if averaged over time,
the diffusion coefficient factorizes in a rigidity and a spatial factor,
energy losses during propagation are negligible, and
escape from the galaxy is the dominant mechanism for CR losses.
Under these conditions, the cosmic-ray spectrum can be written as
$n_\text{CR} (E, \vec x) = n_\odot(E) f(\vec x)$, where $n_\odot(E)$ is the
local CR spectrum, and $f$ is equal to 1 at the position of Earth. A simple
parametrization for the space dependence in this case is given by \cite{lipari}
\begin{equation}
  f(\vec x) = f(r, z) = \frac{\sech (r/R_\text{cr}) \sech (z/Z_\text{cr})}
  {\sech(r_\odot/R_\text{cr})}
\end{equation}
with the hyperbolic secant $\sech x = 1/\cosh x$, $R_\text{cr} = \SI{5.1}{kpc}$
and $Z_\text{cr} = \SI{0.3}{kpc}$. This function is equal to 1 at the position of
the earth $r = r_\odot = \SI{8.5}{kpc}, z = 0$, falls off exponentially at large
distances from the center of the galaxy, and has a vanishing derivative at the
galactic center. Here, $r, z$ are cylindrical coordinates in the
galactic plane with $r = 0, z = 0$ at the galactic center.
As expected, the largest intensity is in the direction of the galactic center,
where the line-of-sight integral
$L(\Omega) = \int_0^\infty f(\vec x_\odot + \tau \hat \Omega) d\tau$ assumes
its maximal value of
\begin{equation}
\begin{split}
  L_\text{max} &= \cosh \frac{r_\odot}{R_\text{cr}} \int_0^\infty \frac{d\tau}
    {\cosh \frac{r_\odot - \tau}{R_\text{cr}}} \\
  &= 2 R_\text{cr} \cosh \frac{r_\odot}{R_\text{cr}} \arctan e^{r_\odot/R_\text{cr}} \\
  &= \SI{38.7}{kpc/sr} = \SI{1.19e23}{cm/sr}.
\end{split}
\end{equation}
The intensity integrated over angular regions corresponding to the inner and outer galaxy,
high latitudes and the whole sky is shown in Fig.~\ref{fig:neutrino_intensity}, together
with the extragalactic flux (see following section).

\subsection{Intensity of boosted neutrinos from extragalactic cosmic rays}
The intensity of diffuse supernova neutrinos boosted by scattering on
extragalactic cosmic rays is given by
\begin{equation}
  \frac{d\phi}{d\epsilon_s} = c \int dz\ 
  \frac{1}{H_0 \sqrt{\Omega_\Lambda + (1 + z)^3 \Omega_m}}
  \dot n_{e,s}(\epsilon_s (1 + z), z),
\end{equation}
where $f(z)$ describes the scaling of the cosmic-ray source density with
redshift $z$, normalized to 1 at $z_\text{min} = \num{2.37e-6}$ corresponding
to the size of the Galaxy, and the emission rate takes into account the
$z$-dependence of the diffuse supernova neutrino background. The extragalactic
emission rate $\dot n_{e,s}$ is given by the same formulae as above, however,
the redshift-dependent cosmic ray flux is instead taken from a simulation of
ultra-high energy cosmic rays done with \textit{PriNCe} \cite{heinze}. The flux
of cosmic rays observed at ultra-high energies on Earth is dominated by nearby
sources due to the short interaction length of cosmic ray nuclei at these
energies; in contrast, the neutrinos produced at high redshifts will propagate
unabsorbed, so that the neutrino flux depends also on the source behaviour at
redshifts that cannot be investigated by fitting the cosmic ray flux observed
at Earth. Since the sources of ultra-high energy cosmic rays are unclear, the
star formation rate, quasar rate \cite{quasar_rate}, and gamma-ray burst rate
\cite{grb_rate} are investigated as possible source densities. The parameters
of the injected sources are given in Table~\ref{tab:extragalactic}. These are
then propagated using the TALYS photodissociation cross-section \cite{talys}
and the Gilmore cosmic infrared background \cite{gilmore}.

\begin{figure}
  \begin{center}
  \includegraphics[width=0.7\textwidth]{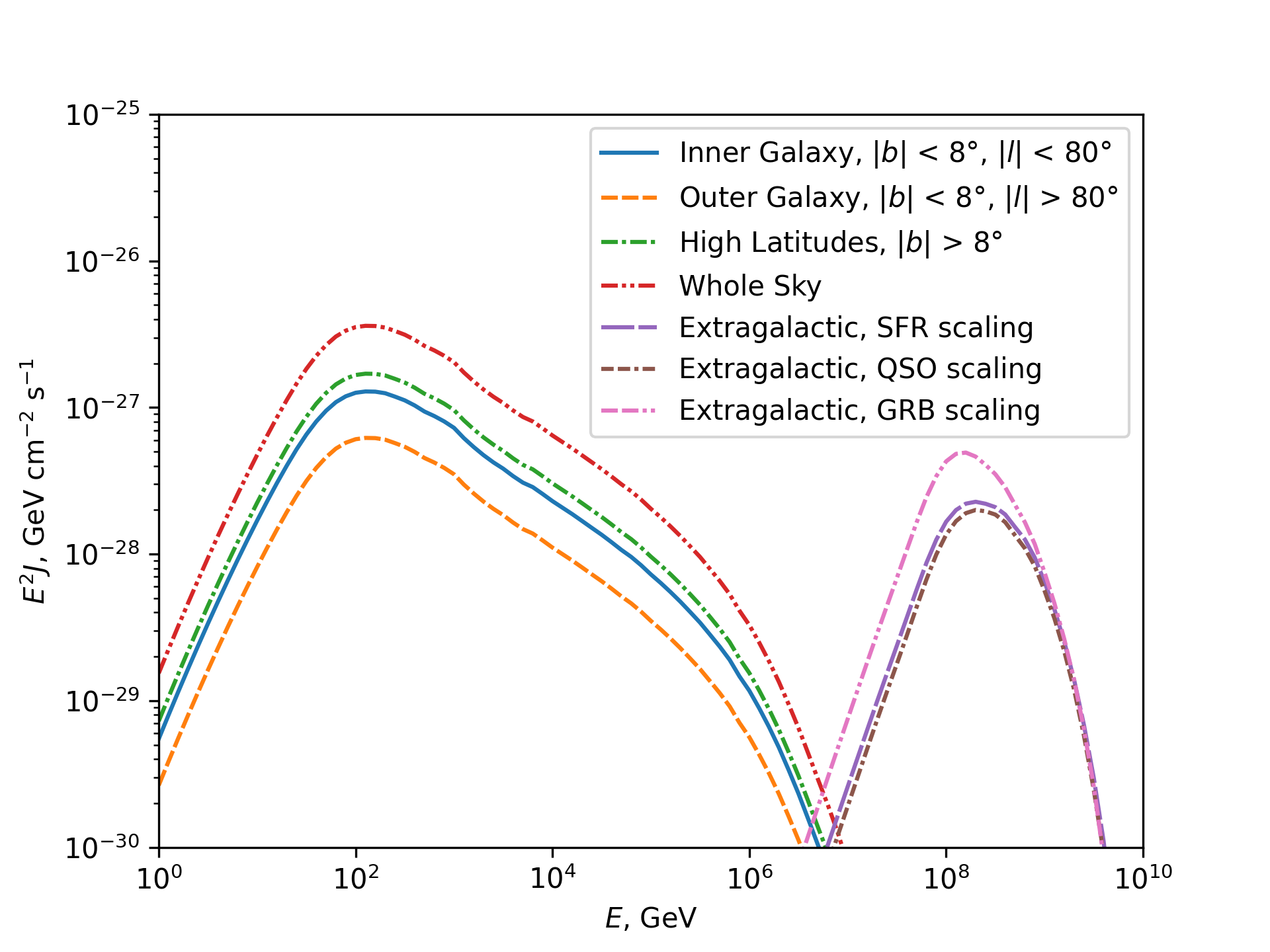}
  \caption{Energy spectra of the CR-boosted diffuse supernova neutrino flux integrated
     over angular regions corresponding to the inner and outer galaxy, high latitudes
     and the whole sky with solid lines; and energy spectra of the extragalactic
     CR-boosted diffuse supernova neutrino flux for different UHECR source density
     scalings with dashed lines.}
  \label{fig:neutrino_intensity}
  \end{center}
\end{figure}

The resulting flux is of similar spectral energy density, but peaks at higher
energies compared to the galactic flux (cf.~Fig.~\ref{fig:neutrino_intensity}).
In contrast to the galactic contribution, this flux is isotropic.

\section{Conclusions}
The cosmic-ray boosted diffuse supernova neutrino background is predicted in
this work. Due to the higher energies of supernova neutrinos compared to relic
neutrinos, not only elastic scattering on nucleons and nuclei but also inelastic
scattering contributes significantly to the boosted flux. Furthermore, the
cosmic ray energies contributing to the observed flux are lower, such that not
only extragalactic ultra-high energy cosmic rays, but also galactic cosmic rays
contribute to the boosted flux.

Similar to the diffuse galactic neutrino emission produced in
cosmic-ray interactions, the emission from scattering on galactic cosmic rays
is concentrated along the galactic plane and follows roughly an $E^{-2.4}$
spectrum. However, the cosmic-ray-boosted diffuse supernova neutrino background
flux is smaller by about 20 orders of magnitude than current models of the
galactic diffuse neutrino emission \cite{cringe}.

The extragalactic component peaks in the region of a few hundred \si{PeV}.
Current experimental limits on the neutrino flux are about 18~orders of
magnitude higher \cite{ehe_prl,ehe_moriond,auger}. Predictions for the as yet
undetected cosmogenic neutrino flux expected from photon-proton interactions
between ultra-high energy cosmic ray protons and the cosmic microwave background
vary depending on assumptions on the composition of the highest-energy cosmic
rays, but even pessimistic calculations predict a much larger flux by more than
10~orders of magnitude \cite{berat_cosmogenic}.

In conclusion, neither the galactic nor the extragalactic component of the
cosmic-ray boosted diffuse supernova neutrino background are expected to be
measurable in the foreseeable future.

\vspace{6pt}

\funding{The author acknowledges funding from the German Bundesministerium für Bildung und Forschung (Förderkennzeichen 05A23PX3).}

\dataavailability{The original contributions presented in this study are included in the article/supplementary material. Further inquiries can be directed to the corresponding author.}

\acknowledgments{The author acknowledges useful discussions with Baobiao Yue, Klaus Helbing, Jiajun Liao and Jiajie Zhang.}

\conflictsofinterest{The author declares no conflicts of interest.} 


\appendixtitles{no} 
\appendixstart
\appendix
\begin{figure}
\centering
\includegraphics[width=0.7\textwidth]{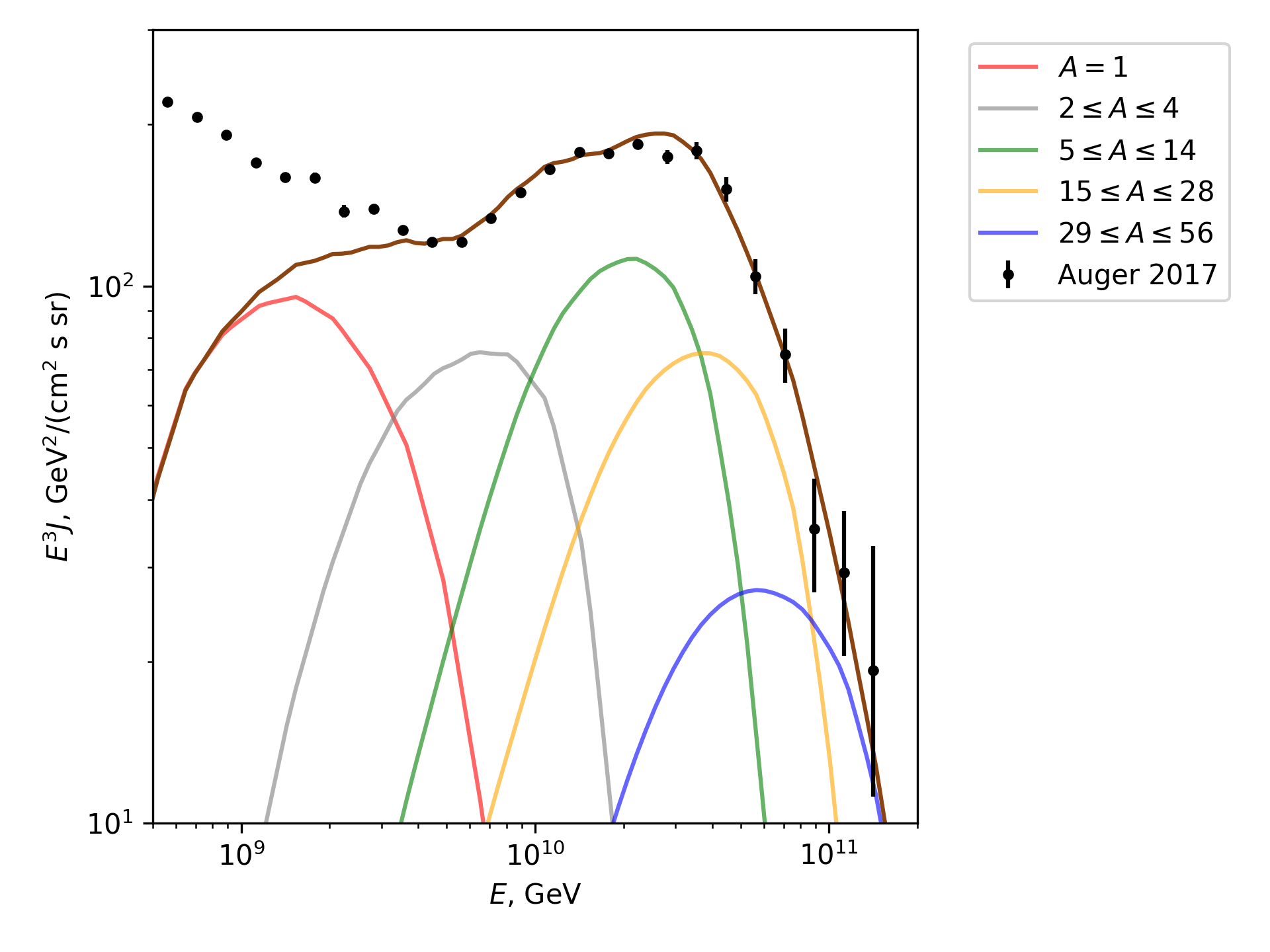}
\caption{Extragalactic cosmic ray flux fit, assuming a source density scaling
as the quasar rate. The fit is to the UHECR measurement data from
\cite{auger2019} above a lower energy of \SI{6e9}{GeV}.}
\label{fig:QSO_UHECR}
\end{figure}

\begin{figure}
\centering
\includegraphics[width=0.7\textwidth]{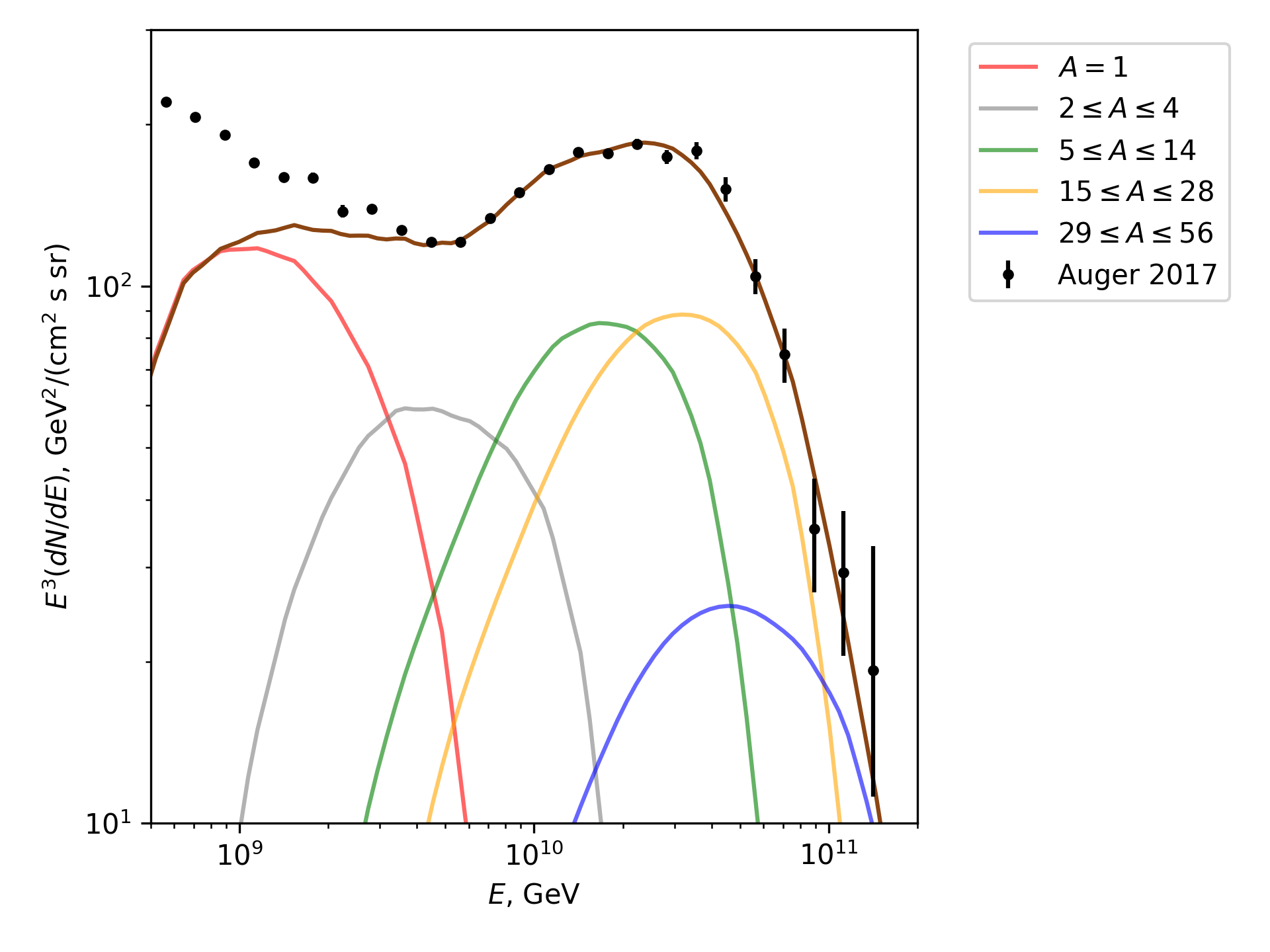}
\caption{Extragalactic cosmic ray flux fit, assuming a source density scaling
as the gamma-ray burst rate. The fit is to the UHECR measurement data from
\cite{auger2019} above a lower energy of \SI{6e9}{GeV}.}
\label{fig:GRB_UHECR}
\end{figure}

\section[\appendixname~\thesection]{Form factors for the elastic
  neutrino-nucleon cross-section according to \cite{ahrens}}
\label{app:neutrino}
The functions $A(Q^2), B(Q^2), C(Q^2)$ in \eqref{eq:ahrens} are given as functions
of the vector form factors $F_1, F_2$ and the axial vector form factor $G_A$ by the
expressions
\begin{align}
  A(Q^2) &= \frac{Q^2}{M_p^2} \left( G_A^2 \left(1 + \frac{Q^2}{4 M_p^2}\right)
    - F_1^2 \left(1 - \frac{Q^2}{4 M_p^2}\right) + F_2^2 \frac{Q^2}{4 M_p^2}
    \left(1 - \frac{Q^2}{4 M_p^2}\right) + F_1 F_2 \frac{Q^2}{M_p^2} \right), \\
  B(Q^2) &= \frac{Q^2}{M_p^2} G_A (F_1 + F_2), \\
  C(Q^2) &= \frac{1}{4} \left(G_A^2 + F_1^2 + F_2^2 \frac{Q^2}{4 M_p^2}\right).
\end{align}
The axial vector formfactor $G_A$ is given by
\begin{equation}
  G_A(Q^2) = \frac{1}{2} \frac{g_A(0)}{(1 + Q^2/M_A^2)^2} (1 + \eta),
\end{equation}
the vector form factors $F_1, F_2$ as linear combination of isovector and isoscalar
form factors
\begin{equation}
  F_1 + F_2 = \alpha G_V^3 + \gamma G_V^0, \quad F_2 = \alpha F_V^3 + \gamma F_V^0
\end{equation}
with the formfactors in the dipole representation given by
\begin{align}
  G_V^3 &= \frac{1}{2} \frac{(1 + \kappa_p - \kappa_n)}{(1 + Q^2/M_V^2)^2}, \\
  G_V^0 &= \frac{3}{2} \frac{(1 + \kappa_p + \kappa_p)}{(1 + Q^2/M_V^2)^2}, \\
  F_V^3 &= \frac{1}{2} \frac{(\kappa_p - \kappa_n)}{(1 + \tau) (1 + Q^2/M_V^2)^2}, \\
  F_V^0 &= \frac{3}{2} \frac{(\kappa_p + \kappa_n)}{(1 + \tau) (1 + Q^2/M_V^2)^2}, \\
  \tau &= \frac{Q^2}{4 M_p^2}
\end{align}
and the coupling constants in the standard model
\begin{align}
  \alpha &= 1 - 2 \sin^2 \theta_W, & \beta &= 1, \\
  \gamma &= -\frac{2}{3} \sin^2 \theta_W, & \delta = 0.
\end{align}
Here, $\theta_W$ is the weak mixing angle, $\kappa_p = \num{1.793},
\kappa_n = \num{-1.913}$ are the anomalous magnetic moments of the proton and neutron,
$M_V = \SI{0.84}{GeV/c^2}, M_A = \SI{1.032}{GeV/c^2}$ are the vector and axial-vector
dipole mass, $g_A(0) = \num{1.26}$, and $\eta = \num{0.12}$.

\begin{adjustwidth}{-\extralength}{0cm}

\reftitle{References}


\bibliography{supernova.bib}

%


\PublishersNote{}
\end{adjustwidth}
\end{document}